\documentstyle{mn}

\input epsf
\newif\ifAMStwofonts

\newcommand{\mum}{$\,\mu$m}

\newcommand{\arcm}{\,arcmin}
\newcommand{\arcs}{\,arcsec}

\ifoldfss
  \ifCUPmtlplainloaded \else
    \NewTextAlphabet{textbfit} {cmbxti10} {}
    \NewTextAlphabet{textbfss} {cmssbx10} {}
    \NewMathAlphabet{mathbfit} {cmbxti10} {} 
    \NewMathAlphabet{mathbfss} {cmssbx10} {} 
  \fi
  \ifAMStwofonts
    \ifCUPmtlplainloaded \else
      \NewSymbolFont{upmath} {eurm10}
      \NewSymbolFont{AMSa} {msam10}
      \NewMathSymbol{\upi}     {0}{upmath}{19}
      \NewMathSymbol{\umu}     {0}{upmath}{16}
      \NewMathSymbol{\upartial}{0}{upmath}{40}
      \NewMathSymbol{\leqslant}{3}{AMSa}{36}
      \NewMathSymbol{\geqslant}{3}{AMSa}{3E}

    \fi
  \fi
\fi 

\ifnfssone
  \newmathalphabet{\mathit}
  \addtoversion{normal}{\mathit}{cmr}{m}{it}
  \addtoversion{bold}{\mathit}{cmr}{bx}{it}
  \newmathalphabet{\mathbfit} 
  \addtoversion{normal}{\mathbfit}{cmr}{bx}{it}
  \addtoversion{bold}{\mathbfit}{cmr}{bx}{it}
  \newmathalphabet{\mathbfss} 
  \addtoversion{normal}{\mathbfss}{cmss}{bx}{n}
  \addtoversion{bold}{\mathbfss}{cmss}{bx}{n}
  \ifAMStwofonts
    \ifCUPmtlplainloaded \else
      \UseAMStwoboldmath
      \makeatletter
      \new@mathgroup\upmath@group
      \define@mathgroup\mv@normal\upmath@group{eur}{m}{n}
      \define@mathgroup\mv@bold\upmath@group{eur}{b}{n}
      \edef\UPM{\hexnumber\upmath@group}
      \new@mathgroup\amsa@group
      \define@mathgroup\mv@normal\amsa@group{msa}{m}{n}
      \define@mathgroup\mv@bold\amsa@group{msa}{m}{n}
      \edef\AMSa{\hexnumber\amsa@group}
      \makeatother
      \mathchardef\upi="0\UPM19
      \mathchardef\umu="0\UPM16
      \mathchardef\upartial="0\UPM40
      \mathchardef\leqslant="3\AMSa36
      \mathchardef\geqslant="3\AMSa3E
    \fi
  \fi
\fi 

\ifnfsstwo
  \DeclareMathAlphabet{\mathbfit}{OT1}{cmr}{bx}{it}
  \SetMathAlphabet\mathbfit{bold}{OT1}{cmr}{bx}{it}
  \DeclareMathAlphabet{\mathbfss}{OT1}{cmss}{bx}{n}
  \SetMathAlphabet\mathbfss{bold}{OT1}{cmss}{bx}{n}
  \ifAMStwofonts
    \ifCUPmtlplainloaded \else
      \DeclareSymbolFont{UPM}{U}{eur}{m}{n}
      \SetSymbolFont{UPM}{bold}{U}{eur}{b}{n}
      \DeclareSymbolFont{AMSa}{U}{msa}{m}{n}
      \DeclareMathSymbol{\upi}{0}{UPM}{"19}
      \DeclareMathSymbol{\umu}{0}{UPM}{"16}
      \DeclareMathSymbol{\upartial}{0}{UPM}{"40}
      \DeclareMathSymbol{\leqslant}{3}{AMSa}{"36}
      \DeclareMathSymbol{\geqslant}{3}{AMSa}{"3E}
    \fi
  \fi
\fi 

\ifCUPmtlplainloaded \else
  \ifAMStwofonts \else 
    \def\upi{\pi}
    \def\umu{\mu}
    \def\upartial{\partial}
  \fi
\fi

\title[A SCUBA Scan-map of the HDF]{A SCUBA Scan-map of the HDF: Measuring the bright end of the sub-mm source counts}
\author[Borys et al.]
  {Colin Borys$^1$, Scott C. Chapman$^2$,
   Mark Halpern$^1$, Douglas Scott$^1$\\
   $^1$Department of Physics and Astronomy,
   University of British Columbia,
   Vancouver BC CANADA V6T 1Z1\\
   $^2$California Institute of Technology,
   Pasadena CA USA 91125
}

\date{2001 July 24}

\pagerange{\pageref{firstpage}--\pageref{lastpage}}
\pubyear{2001}

\begin{document}
\maketitle
\label{firstpage}

\begin{abstract}
Using the 850\mum\ SCUBA camera on the JCMT and a scanning technique 
different from other sub-mm surveys, we have obtained
a 125 square arcminute map centered on the Hubble Deep Field.  
The one-sigma sensitivity to 
point sources is roughly 3\,mJy and thus our map probes the brighter end of
the sub-mm source counts. We find 6 sources with a flux greater than about 12\,mJy
($> 4\sigma$) and, after a careful accounting of incompleteness and flux
bias, estimate the integrated density of bright sources 
$N(>12\,\rm{mJy})= 164^{+77}_{-58}\,\rm{degree}^{-2}$ 
(68 per cent confidence bounds). 

\end{abstract}


\section{Introduction}
Detections of submillimetre emission from high redshift galaxies
highlight the importance of dust in the early history of galaxy
formation.  Observations using the Submillimetre Common
User Bolometer Array (Holland et al.~1999)  of small fields down to the
confusion limit allow us to estimate the
source counts of 850\mum\ sources below 10\,mJy.  In order to learn
more about source counts (and hence to constrain models), the next step
is to search for brighter $(> 10\,$mJy) objects over somewhat larger 
$(> 100\,$square arcminute) fields.

The population of bright sub-mm sources is not well understood.
Current models (Fall, Charlot \& Pei~1996; Blain \& Longair~1996; 
Blain~1998; Eales et al.~2000; Rowan-Robinson~2001)
for source counts in the sub-mm have
been able to account for the observed sources by invoking evolution
which follows the $(1+z)^3$ form required to account for IRAS galaxies
at 60$\,\mu$m, and the powerful radio-galaxies and quasars
(Dunlop \& Peacock~1990).
Euclidean models with no evolution have a
slope of roughly $-1.5$ (i.e. $N(>S)\,{\propto}\,S^{-3/2}$), which cannot
possibly account for the lack of sources observed. With reasonable evolution
(in IRAS-motivated models), the counts steepen sharply at the 10's of
mJy level to roughly $S^{-2.5}$. Given the additional
constraint of not overproducing the sub-mm background, the counts of
relatively weak sources lead to little variation between the models.
However, at the $10$--$30\,$mJy brightness level various evolutionary
models (e.g. Guiderdoni et al.~1998)  show more parameter dependence. 
Moreover, given that
brighter sources are easier to follow up at other wavelengths, their
investigation may help understand the composition of galaxy types and 
their evolution.

In this paper, we present a new estimate for the 
number density of sources which are brighter than $12\,$mJy at 850\mum.
A general discussion of how sources are detected is presented in the 
next section.  To convert the
detections into a source count requires a careful study of the flux bias
and incompleteness, which are unavoidable due to confusion and non-linear
thresholding.  Section three describes Monte-Carlo studies used to 
estimate and correct for these effects and provide a proper number 
count estimate. Finally we discuss correlations between the SCUBA 
scan-map and other sub-mm data sets.  Comparison with data from other
wavelengths is reserved for a longer paper in preparation 
(Borys et al.~2002).

\section{Data collection and analysis}
The SCUBA detector is a hexagonal array of bolometers with a field of 
view of about 2\arcm.  Using a dichroic filter, the radiation is
separated into 850 and 450\mum\ bands which are directed toward 37 and 
91 element arrays respectively. To sample a large region of the sky, 
the telescope scans this array along one of three special directions
with respect to the array, chosen to ensure that sky is fully 
sampled.  The secondary mirror is chopped at a frequency of $7\,$Hz
between a `source' and `reference' position in order to remove
fluctuations due to atmospheric emission.

The standard SCUBA scan-map observing strategy uses multiple
chop throws in two fixed directions on the sky and an FFT-based
deconvolution technique.  This may be appropriate for
regions where structure appears on all scales, but it is not optimal 
for finding point sources. The off-beam pattern gets diluted (making 
it more challenging to isolate faint sources) and the map noise 
properties are difficult to understand.  Our approach is to use a
single chop direction and throw, fixed in orientation on the sky. The
result is a map that has, for each source, a negative `echo' of equal
amplitude in a predictable position.  Although the chop was fixed, we
scanned in all three directions in order to better distinguish signal 
from instrumental noise and sky fluctuations.

A total of 61 scans were obtained in 3 separate runs between 1998 and
2000.  The shape of the region mapped was a square centred on the HDF 
and was
oriented along lines of constant right ascension and declination. 
Using SURF (Jenness \& Lightfoot~1998) and our own custom
software, we are able to isolate and remove large scale
features in the maps and estimate the per-pixel noise level by a
careful accounting of the per-bolometer noise and the frequency with
which each particular pixel is sampled.

\subsection{Scan mapping versus Jiggle map mosaics for large surveys}
Most extra-galactic sub-mm surveys previously done use a series of 
jiggle maps to fully sample a large area of the sky. To ensure uniform
noise coverage, the number of integrations on a particular jiggle-map
field has to be adjusted according to the weather at the time of
observation.  The noise level across a scan-map composed of individual 
scans is very uniform, since each pixel is sampled by many different 
bolometers, whose noise is constant throughout a scan. One advantage 
of jiggle-mapping is that it is the most used SCUBA observing mode and 
thus is already well understood.
Also, it is known that the observing efficiency, $\epsilon$ 
(the ratio of time spent collecting data to elapsed real time) for the 
jiggle-map mode is higher than for scan-mapping, but this is a slightly 
misleading measure for reasons discussed below.

In section 2.3 we describe how we measure fluxes of point sources by
fitting the beam pattern to the map. We denote the reduction in noise 
by fitting the beam pattern, as opposed to fitting the positive beam 
alone, by $\xi$. In the case of scan-mapping, where the beam weights 
are $+1$ and $-1$, we have $\xi_S=\sqrt{2}$.  The beam weights for 
jiggle-mapping, $-0.5,+1,-0.5$ lead to $\xi_J=\sqrt{3/2}$.  We have 
determined that the observing efficiencies are 
$\epsilon_{J} \simeq 0.77$ and 
$\epsilon_{S} \simeq 0.52$ for jiggle and scan-mapping respectively. The
nature of the loss of efficiency in both modes is related to the time
needed to read out data after each exposure.  In the scan-map mode,
the sky is sampled 8 times more often than in jiggle-mode, and hence
the readout time is longer.
 
The overall sensitivity to point sources is 
$\propto \xi\epsilon^{-0.5}$ and, given the numbers above, 
the jiggle-map and scan-map mode are found to be almost equally  
sensitive.

The advantage of using scan-mapping over jiggle-maps in studies of the 
clustering of SCUBA sources is that the latter mode is insensitive to 
scales larger than the array size. This, coupled with the uniform noise 
level and nearly equivalent effective sensitivity should make 
scan-mapping the preferred mode when planning large field surveys.

\subsection{Pointing and flux calibration}

Reliably determining the pointing of scan-maps containing no bright 
sources is problematic.
To check for evidence of systematic pointing errors in our data, we 
re-analysed the Barger et al. (2000) HDF/radio fields, which are a set 
of smaller maps taken within the  HDF flanking field. These maps were 
taken in the well-used jiggle-map mode and were accompanied by several 
pointing checks and calibration measurements.  We compared the maps 
against our scan-map using a nulling test, finding a shift of 
$4\,\pm\,3\,$ \arcs\  to the west. There is no evidence that 
this shift is variable across the field.  The source of this pointing 
offset is not entirely understood, though we note the 3\arcs\ per 
sample scan rate of the telescope and that the shift is along the 
direction of the chop.

The calibration of our map was determined by fitting the beam model to 
observations of standard calibrators taken during the run. 
However, many of the scan-map calibrations turned out to be unusable 
and we had to rely on photometric and jiggle mode calibrations 
obtained during the runs.  Although scan-map flux conversion factors 
differ from those measured in other modes, it is reasonable to assume 
that they are  related by a constant multiplicative factor.  
Hence we were able to derive a relative calibration between nights,
and by comparing the scan-map  against the Barger data, were were able 
to determine the absolute calibration.

\subsection{Source detection}
To construct a model beam-shape we artificially added a very bright 
source into the data and created a map using our analysis pipeline.  
The positive lobe of the beam was Gaussian, but the negative lobe showed 
some evidence of smearing.  This is not surprising considering that 
some of the observations were inadvertently taken using an azimuthal 
chop (4 out of 61 scans), and other observations 
were affected by the `chop-track bug', which caused the off-beams of 
some SCUBA datasets to rotate on the sky even though a fixed coordinate 
frame was requested.

Sources were found by fitting the dual beam pattern, in a least-squares 
sense, to each pixel in the map. Note that this is equivalent to a 
convolution of the map with the beam (Eales et al.~2000) except that it 
also accounts for the pixel by pixel noise.  

A total of six sources were found that have a peak to error-of-fit ratio
greater than 4.  We also rotated the dual-beam pattern by $180^\circ$
and used that as a model; only 3 sources were found, two of
which are associated  with the two brightest sources in the map.  
Monte-Carlo simulations suggest that the other false positive is not 
unexpected.  As an additional check, the 3 runs were compared by eye 
to ensure that each source was evident in all subsets of the data.
An additional 6 sources were recovered with a S/N between 3.5 and 4.0.  
Though more likely to be spurious sources, they are still relatively 
bright and are included to facilitate comparison with data at other 
wavelengths.  A complete list of sources is given in Table~1.

The map shown in Figure~1 is the output from the source fitting 
algorithm.  It must be pointed out that by convolving with the beam, 
the dual-beam pattern is converted into a triple beam. The positive 
source now has two negative echos of half the amplitude spaced by the 
chop throw on either side.

\begin{figure*}
 \epsfxsize=6.5in
\center\epsfbox{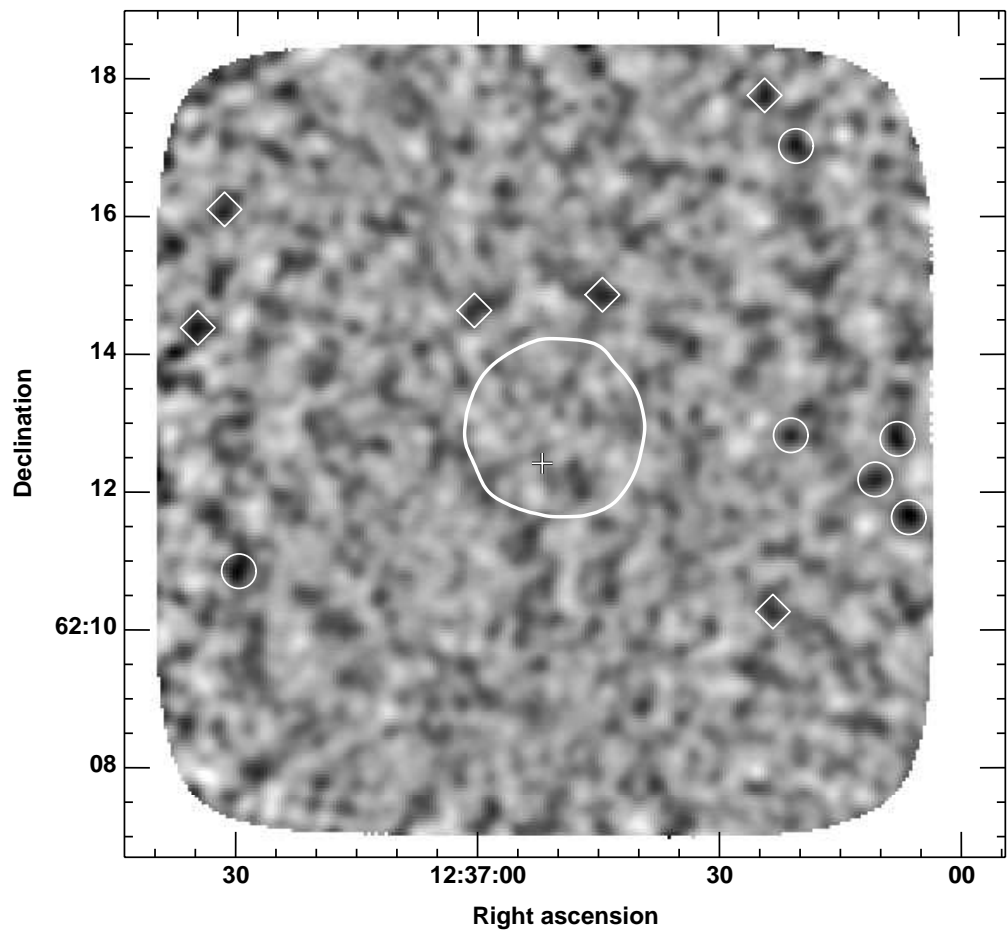} 
 \caption{The 850\mum\ HDF scan-map.  Black corresponds to positive
flux density.
The map size is approximately $11.25 \times 11.25$ arcminutes.  
The chop is 40 arcseconds and is roughly east-west.  The circles 
outline the six $>4\sigma$ sources detected in our survey and the 
diamonds the 6 addition detections above $3.5\sigma$.  The region covered by
Hughes et al. (1998) is denoted by the white outline near the centre of the map,
with their strongest detection, HDF850.1 indicated by the white cross. 
}
 \label{fig:hdf850}
\end{figure*}

\setcounter{table}{0}
\begin{table*}
 \begin{minipage}{115mm}
 \caption{850\mum\ detections in the HDF scan-map.  The top half of the
table lists the $>4\sigma$ sources in order of increasing RA.  The last
six entries are the weaker detections, again ordered according to RA.}
 \label{tab1}
 \begin{tabular}{lllll}
ID & RA(2000) & DEC(2000) & $S_{850}$ (mJy) & S/N \\
HDFSMM$-$3606+1138 & 12:36:06.4  &   62:11:38 & $15.4\pm3.4$ &   4.5 \\
HDFSMM$-$3608+1246 & 12:36:07.8  &   62:12:46 & $13.8\pm3.3$ &   4.2 \\
HDFSMM$-$3611+1211 & 12:36:10.6  &   62:12:11 & $12.2\pm3.0$ &   4.0 \\
HDFSMM$-$3620+1701 & 12:36:20.3  &   62:17:01 & $13.2\pm2.9$ &   4.6 \\
HDFSMM$-$3621+1250 & 12:36:21.1  &   62:12:50 & $11.4\pm2.8$ &   4.0 \\
HDFSMM$-$3730+1051 & 12:37:29.7  &   62:10:51 & $14.3\pm3.2$ &   4.5 \\
\\
HDFSMM$-$3623+1016 & 12:36:23.3  &   62:10:16 & $10.3\pm2.9$ &   3.5 \\
HDFSMM$-$3624+1746 & 12:36:24.2  &   62:17:46 & $12.6\pm3.4$ &   3.7 \\
HDFSMM$-$3644+1452 & 12:36:44.5  &   62:14:52 & $11.4\pm2.9$ &   3.9 \\
HDFSMM$-$3700+1438 & 12:37:00.4  &   62:14:38 & $10.1\pm2.9$ &   3.5 \\
HDFSMM$-$3732+1606 & 12:37:31.6  &   62:16:06 & $12.1\pm3.3$ &   3.6 \\
HDFSMM$-$3735+1423 & 12:37:34.9  &   62:14:23 & $13.4\pm3.8$ &   3.6 \\ 
 \end{tabular}
 \end{minipage}
\end{table*}

\section{Source counts}
\subsection{Completeness correction}

In order to estimate the density of sources brighter than some flux
threshold $S^\prime$, $N(>S^\prime)$, we must account for several
anticipated statistical effects on our list of detected sources.
\begin{enumerate}
\renewcommand{\theenumi}{(\roman{enumi})}
\item The threshold for source detection, $S_T = m\sigma$
is not uniform across the map due to pixel to pixel noise variation.
\item Due to confusion and detector noise, sources dimmer than 
$S_T$ might be scattered above the detection threshold.
\item Similarly, sources brighter than $S_T$ might be missed.
\end{enumerate}

Item (\rm{iii}) is simply the completeness of our list of sources,
and must take into account edge effects and possible source overlaps.
For a source 
density $N(S)$ which falls with increasing flux, item (\rm{ii}) 
typically exceeds item (\rm{iii}), resulting in Eddington bias. 
Thus measured fluxes tend to be brighter than the true fluxes of the 
sources. 

Using Monte Carlo simulations, these  two effects are quantified by 
measuring $f_m(S)$, the fraction of sources at flux $S$ which we would 
detect above the threshold $S_T = m\sigma $. Artificial sources with a 
range of known fluxes are added to the time
series corresponding to random locations (but constrained not
to overlap each other) and run through our source detection pipeline.

One can calculate the ratio of the integrated source count to the number
of sources we detect, 
\begin{equation}
   \gamma(S^\prime)= {{{\int_{S^\prime}^{\infty} {N(S)dS}}}\over{{\int_{0}^{\infty} f_{m}(S)N(S)dS}}}
\end{equation}
where $N(S)dS$ is the number of sources between a flux $S$ and $S+dS$.
Note that $f_{m}(S)$ ranges between zero to unity but 
$\gamma(S^\prime)$ can be larger than one depending on the form of 
$N(S)$ and choice of $S^\prime$.

The calculation of $\gamma(S^\prime)$ from the Monte Carlo estimates of
$f_{m}(S)$ requires a model of the source counts.  We employ the two 
power-law phenomenological form of Scott and White (1999),
\begin{equation}
   N(>S) = {N_0}{{\left({S}\over{S_0}\right)}^{-\alpha}}{{\left(1-{{S}\over{S_0}}\right)}^{-\beta}}.
\end{equation}
and use $S_0=10\,$mJy, $\alpha=0.8$, $\beta=2.5$, and
$N_0=1.55\times10^4\,{\rm deg}^{-2}$.

Obviously the factor $\gamma$ is influenced by what model is used in
the calculation, and other forms of the source spectrum are found in
the literature.  We find $\gamma$ varies by no more than 10 per cent
across a wide range of reasonable parameter values. 

This calculation combines all three effects listed above.  In the
present case,  effects (\rm{ii}) and (\rm{iii}) balance out almost 
exactly.
We calculate $\gamma(12\,$mJy$)=0.95$, and therefore obtain
$N(>12\,$mJy$)=164^{+77}_{-58}\,{\rm deg}^{-2}$ (68 per cent confidence 
limit) based on the six $4\sigma$ sources found in the 125 square 
arcminute map.  
This value is plotted along with estimates from other studies in Figure~2.

\subsection{Flux bias correction}
Our Monte-Carlo studies confirm the results of Eales et al. (2000)
that detected sources tend to be Eddington biased upwards in flux.
At the $4\sigma$ level used in this work, our simulations
indicate that the boost is 20 per cent in apparent flux; Eales et
al. estimate a 40 per cent boost for their $3\sigma$ sources.  Note 
there is no need to use this factor to adjust our estimate of the 
source counts because it is inherently taken care of in the Monte 
Carlo simulation and the counts-weighting; We could equivalently quote
a somewhat higher source density at 10\,mJy.

\begin{figure}
 \epsfxsize=18pc
 \center\epsfbox{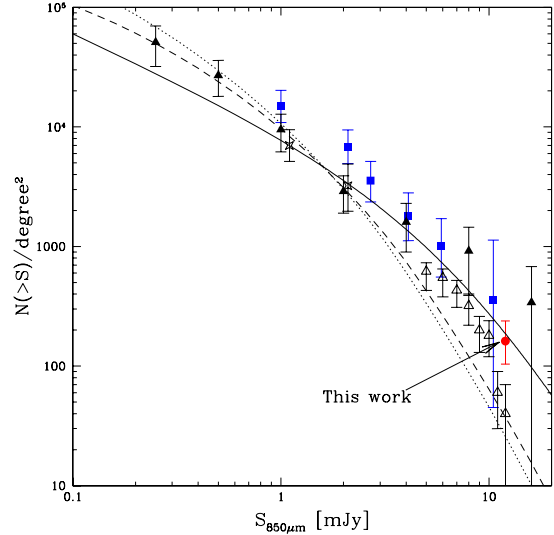} 
 \caption{The 850\mum\ source counts.  The solid circle shows the 
result from the current work.
Counts derived from cluster studies by Chapman et al. (2001) and 
Blain et al. (1999) are shown by the solid squares and triangles 
respectively. The UK\,8 mJy survey counts (Scott et al. 2001) 
are shown as 
open triangles.  Crosses represent the counts from Hughes et al. (1998).
Some points are slightly offset along the flux axis for clarity.
Overlaid is the two power-law model of Scott \& White (solid line), and 
two predictions based on the models of Rowan-Robinson (2001).  
The dashed line represents a universe with
$\Omega_M = 1.0$ and $\Omega_\Lambda = 0.0$ while the dotted line is 
$\Omega_M = 0.3$ and $\Omega_\Lambda = 0.7$.}
 \label{fig:N850}
\end{figure}

\section{Discussion}
We have used other available sub-mm data sets in order to assess the 
authenticity of our detections.  The map of the HDF itself by 
Hughes et al. (1998) cannot be used for direct
comparison because their brightest sources are well below the noise level
of this scan-map.  Some of sources in the Barger HDF fields previously 
mentioned are also slightly too faint to be securely detected given 
the sensitivity of our scan-map, but the brighter ones are found.  
The average 850\mum\ flux from the seven sub-mm sources in the original 
Barger catalogue (Barger et al.~2000) is $8.4\pm0.6\,$mJy in our 
re-analysis of those data.  The average flux from the 
scan-map at the same seven positions is $8.1\pm1.1\,$mJy.  

Perhaps the most striking feature of the scan-map is the group of three 
sources
on the eastern edge (the first three sources in Table 1).  Since the
group fits within a single jiggle-map field, we obtained time via the
CANSERV program to confirm the existence of these three sources.
HDFSMM$-$3606+1138  and HDFSMM$-$3611+1211 differ by no more than 
$2.5\,$mJy between the scan-map and jiggle-map. HDFSMM$-$3608+1246 is 
fainter by roughly $5\,$mJy but is still reasonably consistent with the 
flux derived from the 
scan-map.

If this trio of sources represent true clustering, then the source count 
derived here may be biased by cosmic variance.  Peacock et al. (2000) 
have found weak evidence of clustering in the $\sim 2 \times 2\,$ 
arcminute map of Hughes et al. (1998).  The UK 8\,mJy survey 
(Scott et al. 2000) which covers over 250 square arcminutes, also shows 
signs of clustering.  However these results are inconclusive, and what 
is needed is a very large $(\sim\,1$ degree) survey.

We have significantly improved statistics on sub-mm source counts at the 
$>10\,$mJy level.
Because these sources are brighter than typical SCUBA detections,
they may be easier to follow up at other wavelengths.
Preliminary analysis already indicates good correlation with $\mu$Jy
radio sources and hard x-ray sources, but little indication of optical 
counterparts (as found in other studies).

Based on the re-reduction of the Barger HDF data, we have also found 
that a careful treatment of the noise map allows us to improve the 
S/N for point sources by about 15\%.  Furthermore, we find that 
scan-mapping is an efficient method for making large  maps for
cosmological studies.  Large-scale spatial instrumental effects do not 
appear to be a significant limitation for analysis of such maps.

Our careful data-reduction, confirmation in independent jiggle-maps and 
extensive Monte-Carlo studies lead us to be quite confident in the 
reality of our six $4\sigma$ sources.  We also provide a list of six 
additional sources between $3.5 - 4.0\,\sigma$ to aid in comparison 
with other studies.

\section*{Acknowledgments}
We would like to thank Amy Barger for access to her data before it
was available on the JCMT public archive, and Doug Johnstone for
several useful conversations.  We are also grateful to the staff at the
JCMT, particularly Tim Jenness and Wayne Holland who provided invaluable
advice.
This work was supported by the Natural Sciences and Engineering Research
Council of Canada. The James Clerk Maxwell Telescope is operated by The 
Joint Astronomy Centre on behalf of the Particle Physics and Astronomy 
Research Council of the United Kingdom, the Netherlands Organisation 
for Scientific Research, and the National Research Council of Canada.

\bsp

\label{lastpage}

\end{document}